\documentclass[runningheads]{llncs}
\usepackage[T1]{fontenc}
\usepackage{graphicx}

\usepackage{amsmath}
\usepackage{bm}
\usepackage{bbm, dsfont}

\usepackage{algorithm}
\usepackage{algpseudocode}
\usepackage{enumitem}
\usepackage{todonotes}

\usepackage{booktabs}
\usepackage{xcolor}
\usepackage{array}
\usepackage{tikz}
\usepackage{xspace}
\usepackage[subtle]{savetrees}

\usepackage[caption=false]{subfig}

\newcommand{\mypar}[1]{\vspace{0.1cm}\noindent\textbf{#1.}\xspace}

\begin{document}
\title{How stealthy is stealthy? Studying the Efficacy of Black-Box Adversarial Attacks in the Real World}
\titlerunning{How stealthy is stealthy?}
%
%

    
    
    

\author{Francesco Panebianco\orcidID{0009-0007-1510-2594} \and
Mario D'Onghia\orcidID{0000-0001-9467-1523} \and
Stefano Zanero\orcidID{0000-0003-4710-5283} \and
Michele Carminati\orcidID{0000-0001-8284-6074}}
\authorrunning{Panebianco et al.}
%
\institute{Politecnico di Milano, Milano Italy\\Dipartimento di Elettronica, Informazione e Bioingegneria\\\email{\{name.surname\}@polimi.it}}
\maketitle              
\begin{abstract}
Deep learning systems, critical in domains like autonomous vehicles, are vulnerable to adversarial examples (crafted inputs designed to mislead classifiers). This study investigates black-box adversarial attacks in computer vision. This is a realistic scenario, where attackers have query-only access to the target model. Three properties are introduced to evaluate attack feasibility: robustness to compression, stealthiness to automatic detection, and stealthiness to human inspection. State-of-the-Art methods tend to prioritize one criterion at the expense of others. We propose ECLIPSE, a novel attack method employing Gaussian blurring on sampled gradients and a local surrogate model. Comprehensive experiments on a public dataset highlight ECLIPSE’s advantages, demonstrating its contribution to the trade-off between the three properties.

\end{abstract}

\section{Introduction}
\label{sec:introduction}

Deep learning models for image classification and object detection are crucial in applications like self-driving vehicles~\cite{balasubramaniam_object_2022}, where they identify vehicles, pedestrians, traffic signs, and obstacles. Detection errors pose significant risks to passengers and others. Many proprietary classifiers and object detection networks are accessible via public Application Programming Interfaces (APIs), increasing their exposure. Examples of such APIs are api4ai~\cite{noauthor-api4ai-nodate} and Clarifai~\cite{noauthor-clarifai-nodate}. Evasion attacks exploit adversarial examples, malicious inputs crafted with noise patterns, to induce misclassification. Early works~\cite{kurakin_adversarial_2016,szegedy_intriguing_2014,carlini_towards_2016} demonstrated the feasibility of such attacks under an unrealistic white-box threat model, where attackers have full knowledge of the target system. Black-box \textit{query-based} attacks~\cite{chen_zoo_2017,al-dujaili_sign_2020,croce_sparse-rs_2022,giulivi2023adversarial,guo_simple_2019,maksym_andriushchenko_square_2019,brendel_decision-based_2018,jianbo_chen_hopskipjumpattack_2020,huichen_li_qeba_2020,thibault_maho_surfree_2020} operate under the more practical constraint of interacting with the model only through remote queries. White-box attacks benefit from gradient information, which guides the optimization process~\cite{szegedy_intriguing_2014,kurakin_adversarial_2016,carlini_towards_2016}. In contrast, black-box attacks address the absence of gradient access in two ways: by estimating gradients from queries when confidence scores are available~\cite{chen_zoo_2017}, or using label-only strategies, which are generally less efficient~\cite{brendel_decision-based_2018,jianbo_chen_hopskipjumpattack_2020,huichen_li_qeba_2020,thibault_maho_surfree_2020}. While label-only attacks are often the only viable option for many AI endpoints, computer vision endpoints frequently disclose top-\(k\) confidence scores~\cite{noauthor-api4ai-nodate,noauthor-clarifai-nodate}, enabling more effective attack strategies. In this work, we explore the limitations and strengths of computer vision evasion attacks relative to their chances to succeed in a real-world scenario. We concentrate on some of the most powerful options in the black-box setting with confidence scores: SimBA~\cite{guo_simple_2019}, SimBA-DCT~\cite{guo_simple_2019}, and the Square Attack~\cite{maksym_andriushchenko_square_2019}.
Prior work has provided definitions of ``deployability"~\cite{giulivi2023adversarial} of an attack in the real world. To the best of our knowledge, no existing work has presented a comprehensive formalization encompassing all characteristics of successful real-world attacks. We introduce a framework consisting of three \textit{effectiveness properties}: \textit{Robustness to Compression} (P1), \textit{Stealthiness to Automatic Detection} (P2), and \textit{Stealthiness to Human Inspection} (P3). The experimental evaluation shows that existing black-box attacks fulfill some effectiveness properties at the expense of others. While SimBA-DCT demonstrates strong performance on P3, its results on P1 and P2 are significantly weaker. Similarly, SimBA performs well on P2 but exhibits poor performance on P1 and P3. In contrast, the Square Attack shows a slight advantage on P1 but fails to achieve satisfactory results on P2 and P3. To address this, we propose \textit{ECLIPSE} (\textbf{E}vasion of \textbf{C}lassifiers with \textbf{L}ocal \textbf{I}ncrease in \textbf{P}ixel \textbf{S}parse \textbf{E}nvironment). The attack resists JPEG compression (Joint Photographic Experts Group) while evading both automatic and human detection. ECLIPSE is a confidence-based black-box evasion attack based on Hill Climbing~\cite{luke_essentials_2013}, a well-known metaheuristic for the optimization of functions lacking a closed form. The algorithm metaphorically "climbs the hill" of the objective function by iteratively improving the best solution found so far. Although ECLIPSE is not the first adversarial attack to rely on this method~\cite{dang_evading_2017,chen_zoo_2017}, its notable effectiveness is attributed to two novel steps integrated within the optimization process. The first step involves applying Gaussian blurring~\cite{digital_image_processing} to the estimated gradients before updating the adversarial example.
A comprehensive evaluation on a public dataset highlights the limitations of state-of-the-art attack methods while demonstrating the superior performance of ECLIPSE in resilience to image compression, detectability, and visual stealthiness.

We summarize the contributions of our research in what follows:
\begin{itemize}
    \item We formalize the real-world feasibility of adversarial examples with three measurable effectiveness properties.

    \item We introduce ECLIPSE, a novel attack designed to achieve a balanced trade-off between all the effectiveness properties.

    \item We evaluate ECLIPSE in terms of effectiveness properties, comparing it with state-of-the-art baselines to identify their advantages and drawbacks.
\end{itemize}

\section{Background and Motivation}
\label{sec:background}
    Classifier \textit{evasion} attacks induce misclassification in victim models by crafting \textit{adversarial examples} — inputs perturbed with specialized noise. These attacks are categorized as \textit{targeted}, where the target label is set, or \textit{untargeted} otherwise. Attackers leverage prior knowledge (e.g., architecture, weights) in white-box attacks and empirical guidance (e.g., input-output behavior) in black-box attacks.    
    Adversarial examples exhibit \textit{transferability}~\cite{szegedy_intriguing_2014,goodfellow_explaining_2014,liu_delving_2016,papernot_transferability_2016,inkawhich_perturbing_2020}, allowing them to fool multiple classifiers. Transfer-based attacks exploit this by crafting adversarial examples using surrogate models for black-box targets, albeit with reduced efficacy compared to native black-box methods~\cite{inkawhich_perturbing_2020}.
    Initial efforts focused on white-box attacks (e.g., FGSM~\cite{goodfellow_explaining_2014}, I-FGSM~\cite{kurakin_adversarial_2016}, Carlini\&Wagner~\cite{carlini_towards_2016}), but the impracticality of assuming full model access has redirected attention to black-box scenarios. Here, the model serves as a remote oracle providing either confidence scores or predicted labels. Confidence-based attacks, leveraging scores for efficient gradient estimation~\cite{chen_zoo_2017}, are practical in domains like computer vision, where services often expose \textit{top-k} confidence scores~\cite{noauthor-api4ai-nodate,noauthor-clarifai-nodate}.
    
    \mypar{Black-Box Attacks with Confidence Scores}
    Guo et al.~\cite{guo_simple_2019} introduced \textit{SimBA} (Simple Black-box Attack), which minimizes the search space using orthonormal basis perturbations of fixed step size. The same work introduces SimBA-DCT, which leverages the same strategy but translates it to the frequency domain. Andriushchenko et al.~\cite{maksym_andriushchenko_square_2019} formalized the \textit{Square Attack}, employing random search with $L_2$ or $L_\infty$ norm constraints. Giulivi et al.~\cite{giulivi2023adversarial} proposed \textit{Adversarial Scratches}, generating deployable adversarial examples via superimposed bezier curves. Among the recently proposed attack methods, SimBA, SimBA-DCT, and the Square Attack are particularly notable due to their widespread popularity and frequent use as baseline approaches in comparative studies.

    \mypar{Defenses}
    Adversarial example research exhibits a persistent cat-and-mouse dynamic typical of information security, with defenses repeatedly proposed and circumvented. While a comprehensive review is beyond this discussion's scope, Ahmed Aldahdooh et al.~\cite{ahmed_aldahdooh_adversarial_2021} provide a detailed overview of adversarial defenses. In real-world applications, preprocessing pipelines can disrupt adversarial perturbations. These pipelines may include defenses explicitly designed to counter such attacks. For example, Byun et al.~\cite{byun_effectiveness_2022} propose Small Noise Defense, which adds minor Gaussian noise to neutralize many black-box adversarial examples. However, even routine and often overlooked preprocessing steps, particularly image compression can significantly degrade the efficacy of adversarial examples.

    \subsection{Motivation}
    While prior work addresses the deployability of adversarial examples~\cite{giulivi2023adversarial,papernot_practical_2017}, a systematic framework for defining attack realism remains absent. Three critical factors can impact attack success: (1) \textit{Image Processing}: Compression methods like JPEG, which reduce file size while preserving perceptible details, can also erase subtle adversarial perturbations. (2) \textit{Adversarial Detection}: Pre-inference detection algorithms can thwart attacks, necessitating minimal queries to avoid alerting the service provider. (3) \textit{Visual Stealthiness}: Prominent adversarial noise increases the risk of detection and mitigation during deployment.

    \mypar{Threat Modeling}  
    We consider an attacker aiming to mislead a computer vision classifier via remotely deployed adversarial examples. The attacker queries the model to obtain \textit{confidence scores}, lacking knowledge of the architecture, weights, or training data. Confidence-based attacks enhance reliability by enabling gradient estimation through inference results~\cite{chen_zoo_2017}. Since confidence score access is common in computer vision systems, this attack scenario is realistic. We evaluate state-of-the-art confidence-based attacks, including \textit{SimBA}~\cite{guo_simple_2019}, \textit{SimBA-DCT}~\cite{guo_simple_2019}, and the \textit{Square Attack}~\cite{maksym_andriushchenko_square_2019}. To ensure consistency, we focus on the $L_\infty$-bound version of the Square Attack, aligning with the $L_\infty$ bounds of the other methods.

\section{Effectiveness properties of adversarial examples}
\label{sec:effectiveness-properties}
    Few works on classifier evasion address the real-world deployability of their proposed solutions. We formalize these considerations as measurable features of images and refer to them as \textit{effectiveness properties}. These properties are organized into three orthogonal dimensions: \textit{Robustness to Compression} (P1); \textit{Stealthiness to Automatic Detection} (P2);  \textit{Stealthiness to Human Inspection} (P3). While some of these properties were individually evaluated in prior work~\cite{giulivi2023adversarial}, to the best of our knowledge we are the first to introduce such formalization of attack effectiveness in the real world. 
    
    \mypar{Robustness to Compression (P1)}
    Images shared on the internet often undergo various processing operations, such as resizing and compression, which are determined by the requirements of the hosting service. Compression, particularly the JPEG format (Joint Photographic Experts Group), is the most prevalent, serving to reduce data transfer and storage costs. As a form of processing, compression can impact the efficacy of adversarial examples to varying degrees. This characteristic was already partially formalized by prior work~\cite{giulivi2023adversarial}, although it was considered the only property to make attacks ``deployable". This property is assessed by measuring the proportion of adversarial examples whose prediction confidence drops below a specified threshold following image compression.
    
    \mypar{Stealthiness to Automatic Detection (P2)}
    Frequency-domain detection has demonstrated high efficacy against state-of-the-art black-box attacks. While ideal for perturbations crafted in the frequency domain (e.g., SimBA-DCT~\cite{guo_simple_2019}), it can also detect some pixel-space attacks, as detailed in Section \ref{sec:experiments}. Remote classification services are accessed via repeated queries, making attack patterns susceptible to stateful defenses~\cite{chen_stateful_2019,esmaeili_iiot_2022,li_blacklight_2022,choi_piha_2023}. While recent methods like OARS~\cite{ryan_feng_stateful_2023} allow black-box attacks to evade such measures, firewalls may still flag anomalous query spikes, and frequent queries can inflate costs.
    
    \mypar{Stealthiness to Human Inspection (P3)}
    Stealthiness to human inspection denotes the ability of adversarial examples to appear indistinguishable from benign inputs to human observers, achieved through minimal perturbations that blend with natural image features. Commonly, the perturbation norm~\cite{guo_simple_2019,maksym_andriushchenko_square_2019,al-dujaili_sign_2020} is minimized to improve stealthiness, though it inadequately captures human perception. A survey by Liu et al.~\cite{liu_hide_2022} highlights the absence of a universal metric for this purpose, necessitating human evaluations for reliable assessments. Accordingly, this property is evaluated by requiring human auditors to assess the extent to which the image appears altered.

\section{ECLIPSE}
\label{sec:eclipse}
    \begin{algorithm}
	\caption{ECLIPSE Algorithm, where $f$ is the remote oracle returning the score of the target class, $H$ is the GradCAM heatmap from the local model, $M$ is the mask, $\tau_t$ is the mask threshold at iteration $t$}
	\begin{algorithmic}[1]
		\Require the original image $x$, $L_\infty$ perturbation budget $\beta$, the maximum number of iterations $I, \epsilon_0$, the sample size $s,$ the gaussian blur kernel size $k$, the gaussian distribution's standard deviation $\sigma, width, height$
		\State Get GradCam heatmap from the local model $H = GradCAM_{local}(x)$
		\State Initialize current best solution $C_0 = x$
		\State Initialize gradient buffers $\nabla[i,j,c] = 0\ \ \ \ \forall i \in [1,height], j \in [1,width], c \in [1,3]$
		\State Initialize mask $M_0[i,j] = 1\ \ \ \ \forall i \in [1,height], j \in [1,width]$
		\State $fitness_0 = f(x)$
		\State Initialize mask threshold $\tau_0 = 0.0$
		\For{$t = 1$ to $I$}
		\State Sample batch $B_s$ of $s$ coordinates $(i,j,c)$ in $M_{t-1}$ without replacement
		\State $\nabla [i,j,c] = f(C_{t-1} + \mathbbm{1}_{(i,j,c)}) - f(C_{t-1})\ \ \ \forall (i,j,c) \in B_s$
		\State Copy gradients to be processed $\delta = \nabla$
		\State $\delta = GaussianBlur(\delta, (k,k), \sigma)$
		\State $A_t = C_{t-1} + \epsilon_{t-1} \frac{\delta}{\max{\text{abs}(\delta)}}$
		\State Clip $A_t$ such that $-\beta \le A_t[i,j,c] - x[i,j,c] \le \beta\ \ \ \forall i \in [1,height], j \in [1,width], c \in [1,3]$
		\State Clip $A_t$ such that $0 \le A_t[i,j,c] \le 1\ \ \ \forall i \in [1,height], j \in [1,width], c \in [1,3]$
		\If{$f(A_t) > fitness_{t-1}$}
			\State $fitness_t = f(A_t)$
			
			\State $\epsilon_{t} = \max \{0.02, 0.95\epsilon_{t-1}\}$
		\Else
			\State $fitness_t = fitness_{t-1}$
		\EndIf
		\State $\tau_t=\min \{0.5, \tau_{t-1}+0.01\}$
		\State $M_t=ThresholdMask(H, \tau_t)$
		\If{$Area(M_t) < min\_area$ \textbf{or} we already sampled $> 0.75Area(M_t)$}
			\State $M_t[i,j]=1\ \ \ \ \forall i \in [1,height], j \in [1,width]$
		\EndIf
		
		\If{$fitness_t > fitness_{t-1}$}
			\State $C_t = A_t$
			\If{$fitness_t > 0.5$}
				\Return $C_t$
			\EndIf
		\EndIf
		\EndFor
		\State Return failure if no solution is found within the max number of iterations
	\end{algorithmic}
 \label{alg:algorithm}
\end{algorithm}

\begin{figure}[t]
        \centering
        \includegraphics[width=0.75\linewidth]{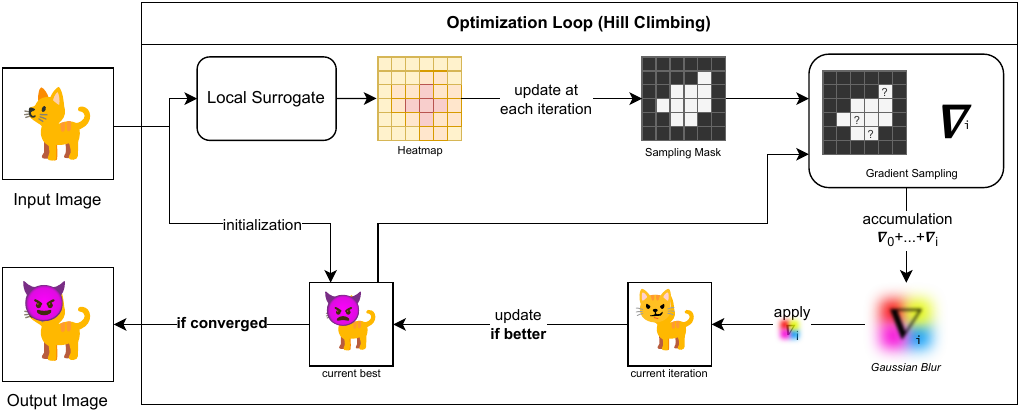}
        \caption{Main steps of the ECLIPSE algorithm: The perturbation mask is computed initially, while other steps iterate until convergence.}
        \label{fig:eclipse-diagram}
    \end{figure}

We propose \textit{ECLIPSE} (\textbf{E}vasion of \textbf{C}lassifiers with \textbf{L}ocal \textbf{I}ncrease in \textbf{P}ixel \textbf{S}parse \textbf{E}nvironment), a targeted evasion attack incorporating two novel techniques aimed at satisfying all three effectiveness properties. Figure~\ref{fig:eclipse-diagram} illustrates how these techniques are embedded in the attack procedure. The first technique involves processing estimated gradients with Gaussian blurring~\cite{digital_image_processing}. The second technique masks the gradient sampling area using information from a local surrogate model. These two components are integrated into an optimization loop that performs \textit{Hill Climbing}~\cite{luke_essentials_2013}, a popular meta-heuristic that iteratively improves the candidate solution, effectively "climbing the hill" of the objective function. The quality of the solution is assessed based on the confidence of the prediction. Consequently, while this approach is effective when confidence scores are available, it is not applicable in label-only settings.

\begin{figure}[t]
        \centering
        \includegraphics[width=0.5\linewidth]{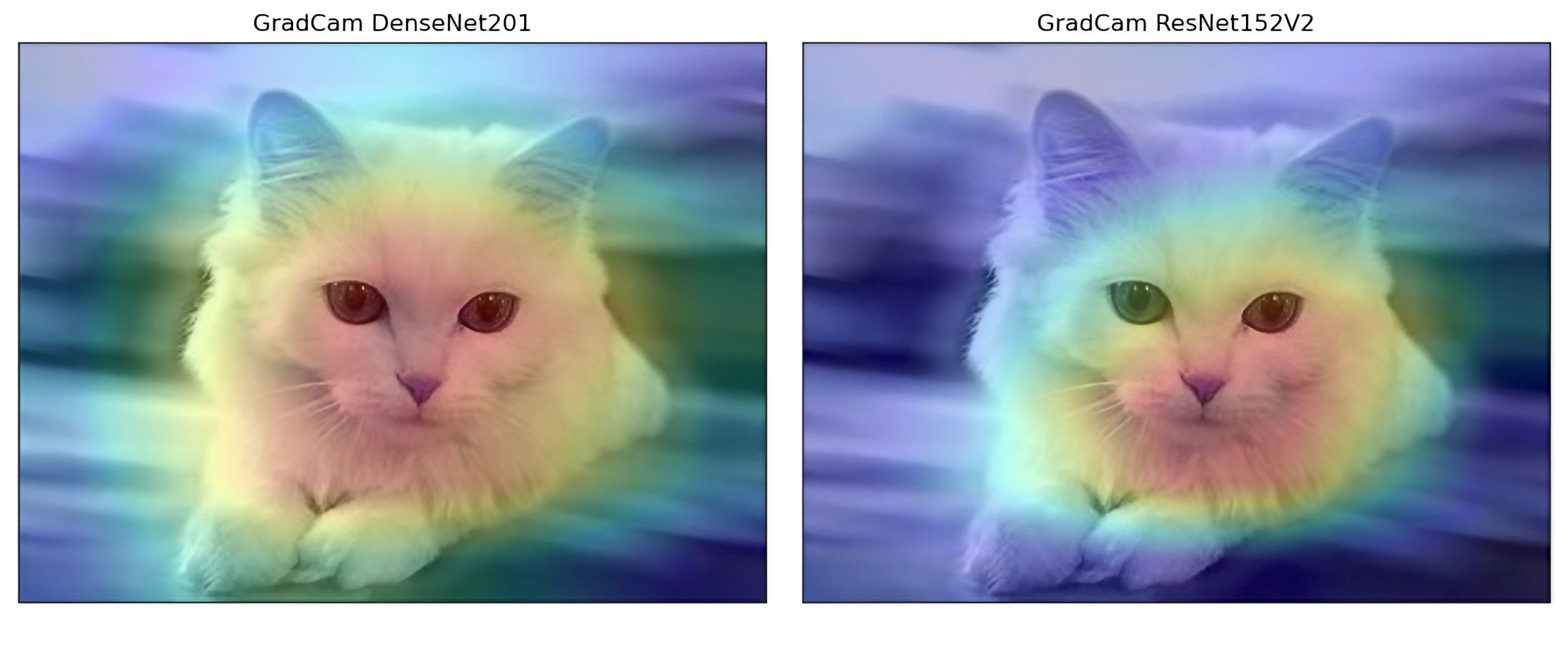}
        \caption{Example comparison of GradCAM heatmaps on a local surrogate and a remote model. Cat image is from the Animals-10~\cite{corrado_alessio_animals-10_nodate} dataset.}
        \label{fig:local-model-role}
    \end{figure}

    \mypar{Local Surrogate}
    Adversarial optimization benefits from identifying relevant input features, which is challenging in query-only scenarios. Surrogate models trained on the same task can approximate remote behavior. We thus leverage a white-box explainability technique (GradCAM~\cite{selvaraju_grad-cam_2017}, Gradient-weighted Class Activation Mapping) to generate saliency maps on the local surrogate. These maps guide perturbations by creating masks that restrict gradient estimation to relevant areas, improving convergence. Figure \ref{fig:local-model-role} shows this concept in practice.
    
    \mypar{Gaussian Blurring}
    Gaussian blurring~\cite{digital_image_processing}, a convolutional process with Gaussian filters, is used in ECLIPSE to smooth estimated gradients rather than image pixels. This approach reduces query count by interpolating sparse gradient values to neighbor coordinates, assuming gradient continuity. While Gaussian blur has been used in training optimizers~\cite{andrew_starnes_gaussian_2023} and defensive techniques~\cite{muzammal_naseer_local_2019}, ECLIPSE uniquely integrates it into gradient-based adversarial attacks.



    \mypar{Attack Parameter Scheduling}
    ECLIPSE employs adaptive scheduling of attack parameters to facilitate convergence. The noise multiplier (\textit{learning rate}) is exponentially decreased to stabilize convergence~\cite{hanneke_rates_2011}. Similarly, the mask threshold is linearly reduced over iterations, focusing perturbations on increasingly relevant areas. These strategies ensure efficient and targeted attack progression.
   
    

\section{Experimental Evaluation}
\label{sec:experiments}
All experiments have been performed on a ResNet152V2 model as the remote oracle. The local surrogate for ECLIPSE experiments is a DenseNet201. It was trained on a subset of ImageNet with 12 classes. Experiments have been run on the Animals-10 dataset~\cite{corrado_alessio_animals-10_nodate}, where images belonging to the ground truth of \textit{cat} are used to generate targeted adversarial examples to be misclassified as \textit{dog}. 

\mypar{Robustness to Processing (P1)}
To assess the robustness of adversarial examples against common image processing during upload, we evaluated JPEG compression, a prevalent internet image format. Table~\ref{tab:auto_comp_attack_parameters} enumerates attack parameters for the experiments. Three metrics were used to evaluate robustness: the median confidence score difference, the percentage of adversarial examples with low confidence loss (below $0.3$), and the percentage of examples that "survived" compression (losing less than $0.05$ confidence, remaining effective or improving). Results, summarized in Table~\ref{tab:compression-comparison}, highlight the varying levels of robustness across the attacks. We can see from this evaluation that ECLIPSE excels in minimizing the loss for the vast majority of adversarial examples it generates. ECLIPSE and the Square Attack $L_\infty$ have a similar survival rate, which is much higher than that of other attacks. In the median case, however, the Square Attack $L_\infty$ performs much worse. Both SimBA and SimBA-DCT fail to produce compression-surviving perturbation, with a very low amount of adversarial examples still effective after processing. The remarkable performance of ECLIPSE, and to a lesser extent that of the Square Attack $L_\infty$, is likely attributable to the coarser perturbation granularity. In fact, the former performs Gaussian blurring, while the latter overlays large squares on the original image.


\begin{table}[b]
    \centering
    \scriptsize
    \setlength{\tabcolsep}{10pt} 
    \caption{Comparison metrics for JPEG compression on different attacks.}
    \begin{tabular}{lrrr}
        \toprule
        Attack & Median Loss & Low-loss\% & Surviving\% \\
        \midrule
        \textbf{ECLIPSE} & \textbf{0.15} & \textbf{89.33} & \textbf{18.67} \\
        SimBA & 0.5 & 4.00 & 2.00 \\
        SimBA-DCT & 0.5 & 4.67 & 1.33 \\
        Square Attack $L_\infty$ & 0.50 & 26.67 & 15.33 \\
        \bottomrule
    \end{tabular}
    \label{tab:compression-comparison}
\end{table}

\begin{figure}[t]
    \centering
    \includegraphics[width=0.8\linewidth]{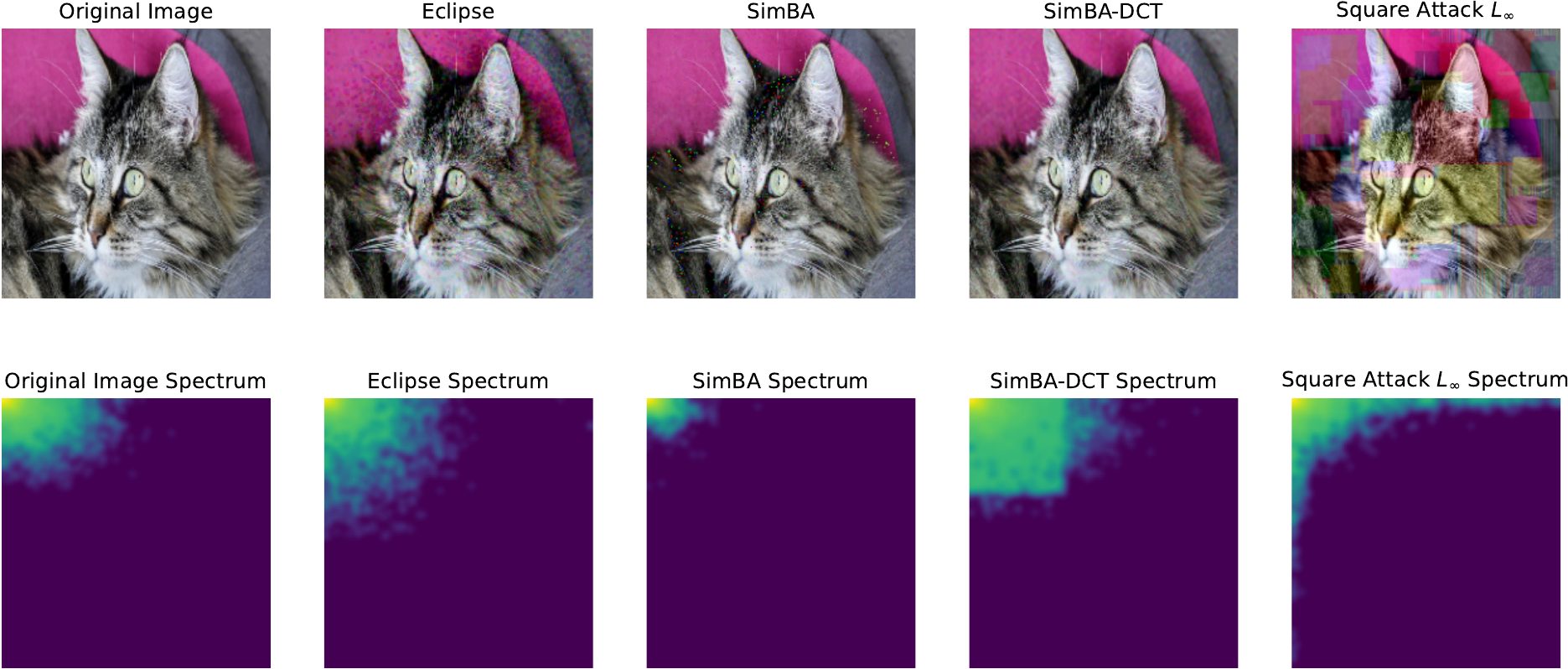}
    \caption{Visual comparison of the processed DCT spectra of adversarial examples generated by each attack against the unaltered image (leftmost).}
    \label{fig:dct-spectra-visual}
\end{figure}

\begin{figure}[t]
    \centering
    \includegraphics[width=0.65\linewidth]{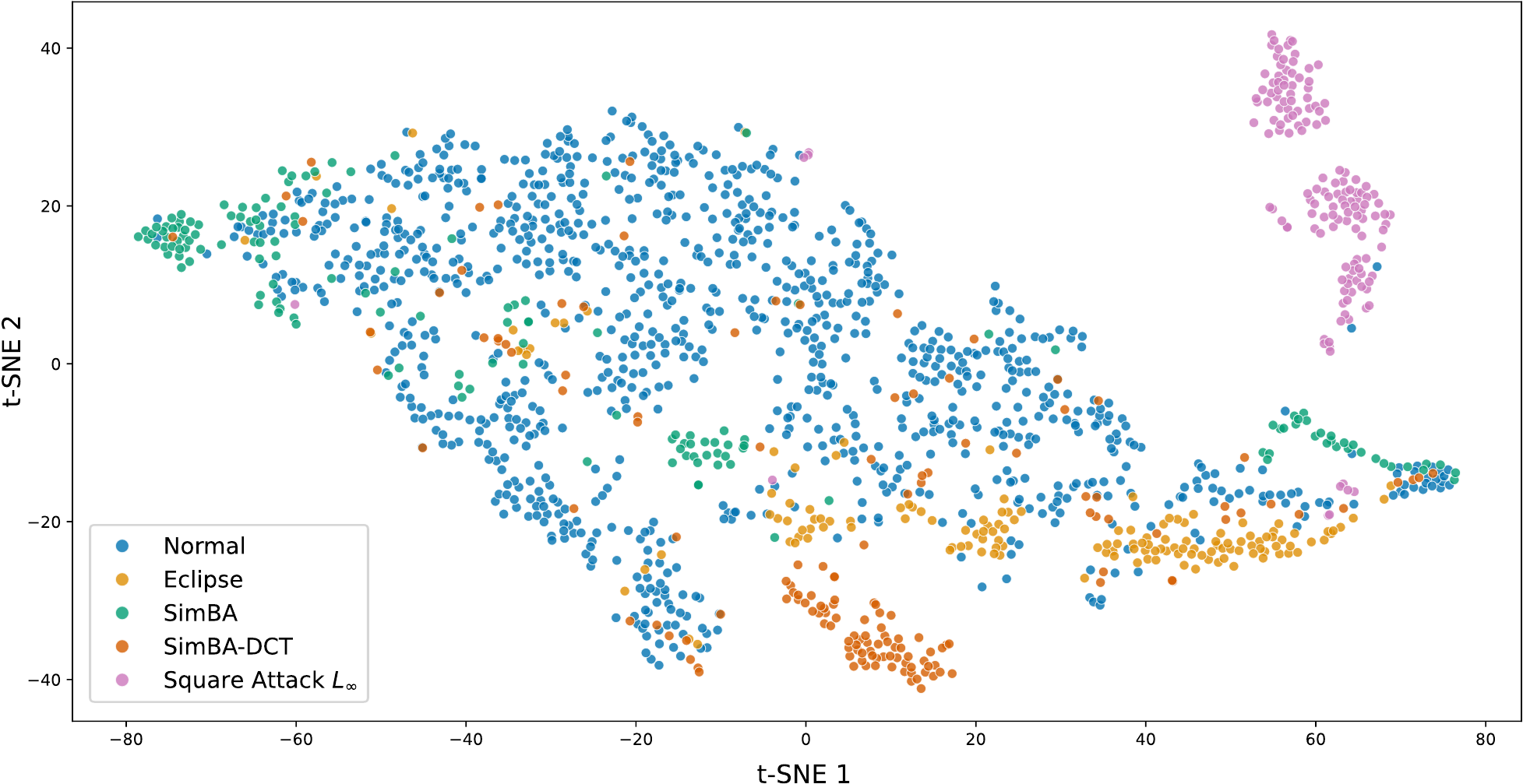}
    \caption{Scatterplot of projected spectral features using t-SNE dimensionality reduction.}
    \label{fig:tsne-dct-spectra}
\end{figure}

\mypar{Stealthiness to Automatic Detection (P2)}
We evaluated adversarial attack strategies on two aspects: detection avoidance and query efficiency. Detection was analyzed using a classifier trained to separate original images from adversarial examples, with features extracted from Discrete Cosine Transform (DCT) spectra. Experiments were conducted under consistent parameters for four attacks (Table \ref{tab:auto_comp_attack_parameters}), using 150 samples per attack. Adversarial attacks introduce distinct spectral changes observable in processed DCT spectra (Figure \ref{fig:dct-spectra-visual}). SimBA-DCT perturbs primarily lower frequencies, creating a square of intensity, while Square Attack $L_\infty$ introduces high-frequency artifacts due to its noise initialization process. We use t-SNE~\cite{tsne_cit} (t-distributed Stochastic Neighbor Embedding) to perform dimensionality reduction. The technique identifies a lower-dimensional manifold that preserves the local structure of the high-dimensional input. This property is particularly useful for exploratory data analysis, as it eases the rapid recognition of data similarities. We observe distinct clusters for Square Attack $L_\infty$ and, partially, for SimBA-DCT (Figure \ref{fig:tsne-dct-spectra}). Binary Support Vector Machines (SVMs) with polynomial kernels were trained to classify adversarial examples. ECLIPSE and SimBA were indistinguishable from benign images, while SimBA-DCT achieved actionable separability (precision = 1.0, recall = 0.47), and Square Attack $L_\infty$ was highly detectable (Area Under the Curve, AUC = 0.96).

\begin{table}[b]
    \centering
    \scriptsize
    \caption{Parameters chosen for comparisons on automatic detection and robustness to processing. $k$ is the Gaussian blurring kernel size. $p$ is the Square Attack's start area ratio. N/A indicates the parameter does not apply to the attack procedure.}
    \begin{tabular}{|c|c|c|c|c|c|}
        \hline
        Attack & Step Size & $L_\infty$ budget & k & p & Max Iterations \\ \hline
        ECLIPSE & 0.1 & 0.1 & 3 & N/A & 1000 \\ \hline
        SimBA & 0.1 & 0.1 & N/A & N/A & 100000 \\ \hline
        SimBA-DCT & 0.1 & 0.1 & N/A & N/A & 100000 \\ \hline
        Square Attack $L_\infty$ & 0.1 & 0.1 & N/A & 0.2 & 10000 \\ \hline
    \end{tabular}
    \label{tab:auto_comp_attack_parameters}
\end{table}

Query efficiency, measured as the number of calls to the victim model, showed that SimBA-DCT and Square Attack $L_\infty$ converge faster than others, but this efficiency is undermined by their high detectability. The Square Attack required fewer queries but failed to converge for 13/150 samples. Despite superior efficiency, its high detectability diminishes practical utility. Our findings highlight the trade-offs between attack efficiency and susceptibility to detection, emphasizing the need for balanced evaluation metrics.

\begin{table}[b]
    \centering
    \scriptsize
    \caption{Cross-validation metrics of binary Support Vector Machine Classifiers to distinguish original images from adversarial examples using processed spectral features. The area under the ROC curve is highlighted in bold.}
    \resizebox{\textwidth}{!}{
    \begin{tabular}{lccccc}
        \hline
        Comparison & Accuracy & Precision & Recall & F1-score & \textbf{ROC AUC} \\
        \hline
        Normal vs ECLIPSE & 0.87 ($\pm$ 0.01) & 0.03 ($\pm$ 0.20) & 0.01 ($\pm$ 0.08) & 0.02 ($\pm$ 0.11) & \textbf{0.50 ($\pm$ 0.03)} \\
        Normal vs SimBA & 0.90 ($\pm$ 0.03) & 0.80 ($\pm$ 0.33) & 0.26 ($\pm$ 0.23) & 0.39 ($\pm$ 0.29) & \textbf{0.63 ($\pm$ 0.12)} \\
        Normal vs SimBA-DCT & 0.93 ($\pm$ 0.03) & 1.00 ($\pm$ 0.00) & 0.47 ($\pm$ 0.25) & 0.63 ($\pm$ 0.24) & \textbf{0.73 ($\pm$ 0.12)} \\
        Normal vs Square Attack $L_\infty$ & 0.99 ($\pm$ 0.02) & 0.99 ($\pm$ 0.05) & 0.91 ($\pm$ 0.17) & 0.95 ($\pm$ 0.10) & \textbf{0.96 ($\pm$ 0.09)} \\
        \hline
    \end{tabular}
    }
    \label{tab:svm-dct-results}
\end{table}

\begin{figure}[t]
    \centering
    \includegraphics[width=0.34\linewidth]{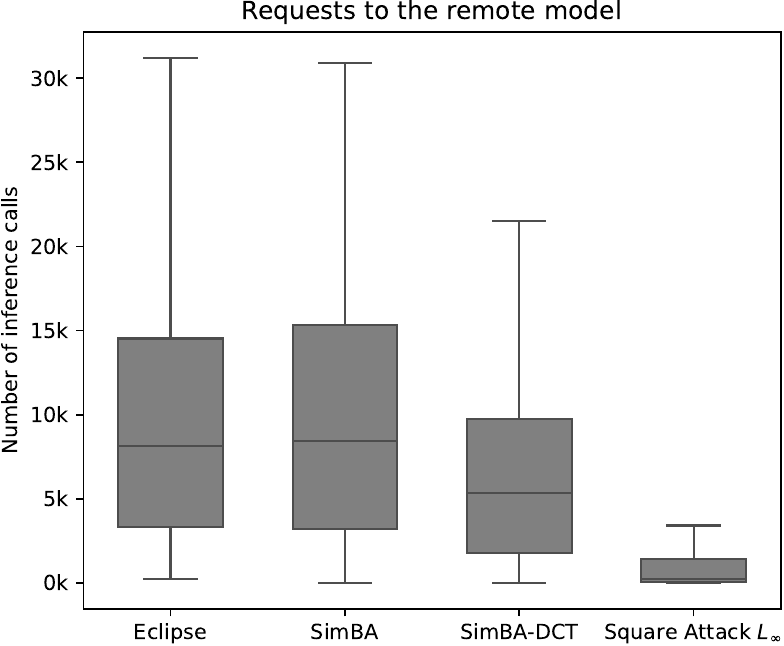}
    \includegraphics[width=0.33\linewidth]{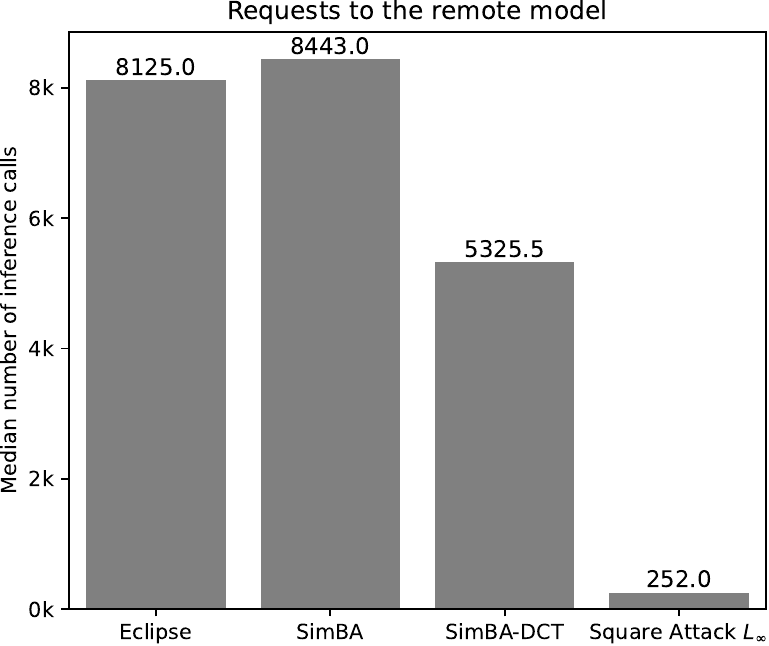}
    \caption{Distribution of requests to the remote model as boxplot without outliers. On the right, the barplot shows the median request count for each attack.}
    \label{fig:infcalls}
\end{figure}

\begin{figure}[t]
    \centering
    \includegraphics[width=0.65\linewidth]{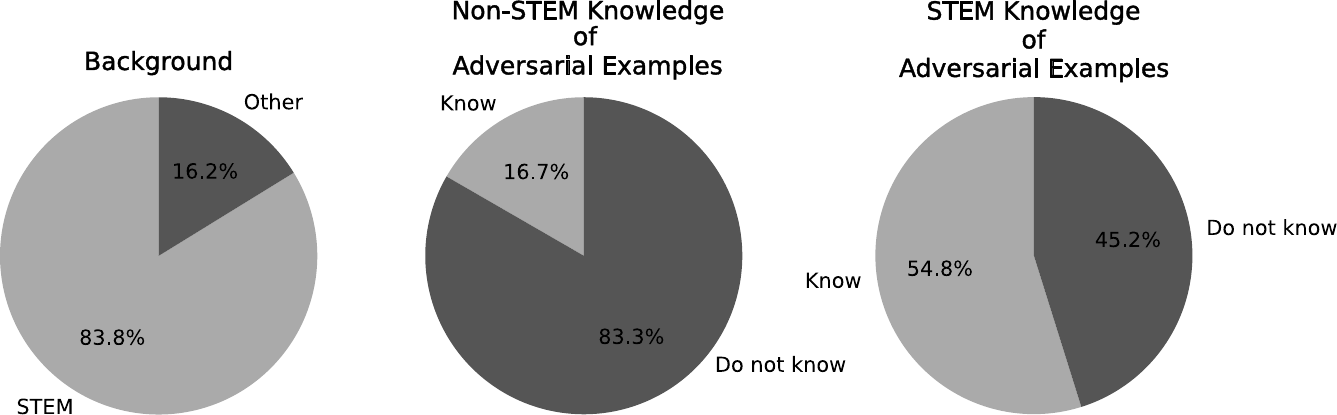}
    \caption{Proportion of respondents from S.T.E.M. fields, awareness of adversarial examples within S.T.E.M. and Non-S.T.E.M. population.}
    \label{fig:participation-background}
\end{figure}

\mypar{Stealthiness to Human Inspection (P3)}
To address the capacity of each attack to transparently blend and avoid human detection, we have administered a survey to 127 people across different backgrounds. Figure \ref{fig:participation-background} shows the distribution of the surveyed population in terms of scientific/non-scientific background. The Ethical statement at the end of this paper discusses the population distribution and other potential concerns. To mitigate bias in perception scores arising from question structure implying malicious edits in certain images, we first uniformly sampled some benign images and then incorporated all corresponding adversarial examples from considered attacks. Each question of the survey showed a picture (either a clean image or an adversarial example) and asked to rate how much they thought the image may have been altered. Valid scores for the answers are integer numbers from 0 to 3, where 0: Not Altered, 1: Slightly Visible, 2: Visible but could fool some people, and 3: Very much visible.

\begin{figure}[t]
    \centering
    \includegraphics[width=0.6\linewidth]{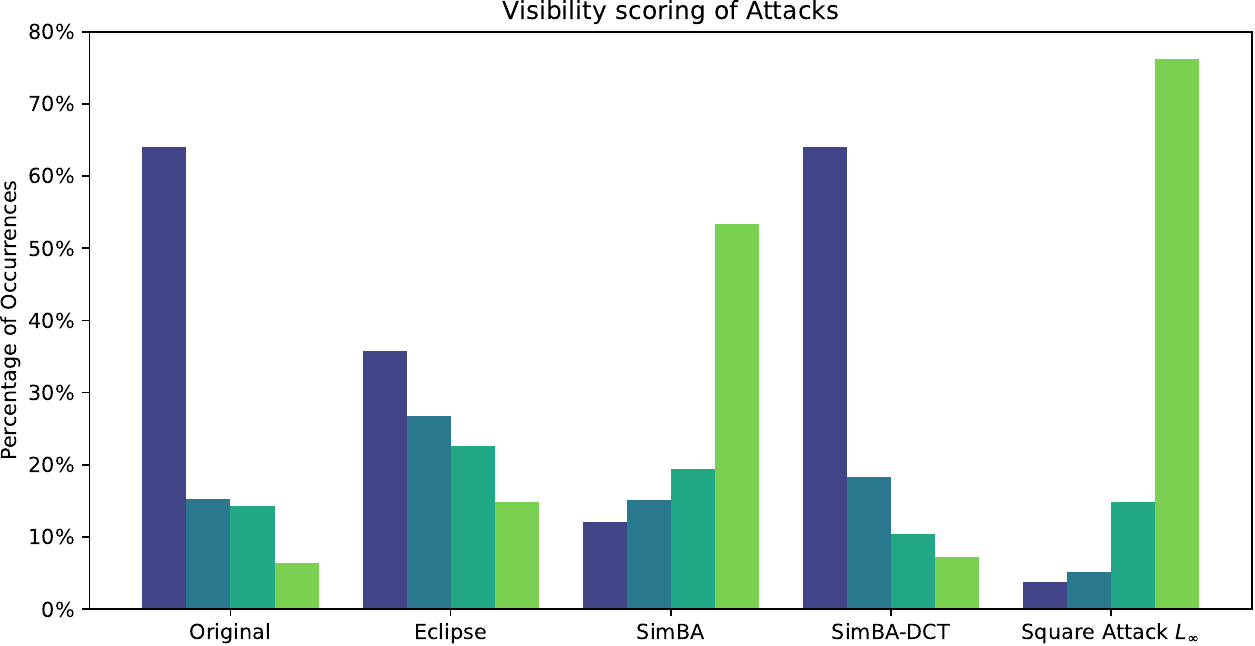}
    \caption{Distribution of visibility scores for each attach on the general population. Scores are from 0 (not visible, blue) to 3 (very much visible, light green). The leftmost category corresponds to unaltered images.}
    \label{fig:visibility-general}
\end{figure}

All images for the study were generated with a perturbation budget of $L_\infty=0.05$. Before showing any picture, the survey asked the participant if they knew what an "adversarial example" is. Figure \ref{fig:participation-background} shows the distribution of people who know the concept in the S.T.E.M. (Science, Technology, Engineering, and Mathematics) and Non-S.T.E.M. groups. Unsurprisingly, the latter seems to have a much smaller percentage of people that are aware of adversarial examples. Comparing the average score given for each attack and to unaltered images, it is apparent that the ranking of the visibility of attacks is the same in both population groups. However, the population that did not know the concept of adversarial examples was observed to be more "paranoid". The general scoring for all images is higher for this group, even for unaltered images. Figure \ref{fig:visibility-general} shows a summary visualization of the scoring given by the general population for each attack and unaltered images. SimBA-DCT can produce very convincing adversarial examples, yielding a score distribution that is very similar to that of the original images. In second place we have ECLIPSE, which is slightly easier to recognize, but remains undetected by a large portion of survey subjects. If we consider both samples that are not recognized (score 0) and slightly visible ones (score 1) we observe a cumulative percentage of $62.55\%$ for ECLIPSE and $82.34\%$ for SimBA-DCT. SimBA and the Square Attack $L_\infty$ follow with undeniably poor results: more than 50\% of the survey participants consider adversarial examples from these attacks to be "very much visible". Their cumulative distribution for scores 0 and 1 is $27.23\%$ and $8.94\%$ respectively.

\begin{figure}[tp]
\centering
\subfloat[Gaussian blurring.\label{fig:eclipse-ablation-kernel}]{%
  \includegraphics[width=0.24\textwidth]{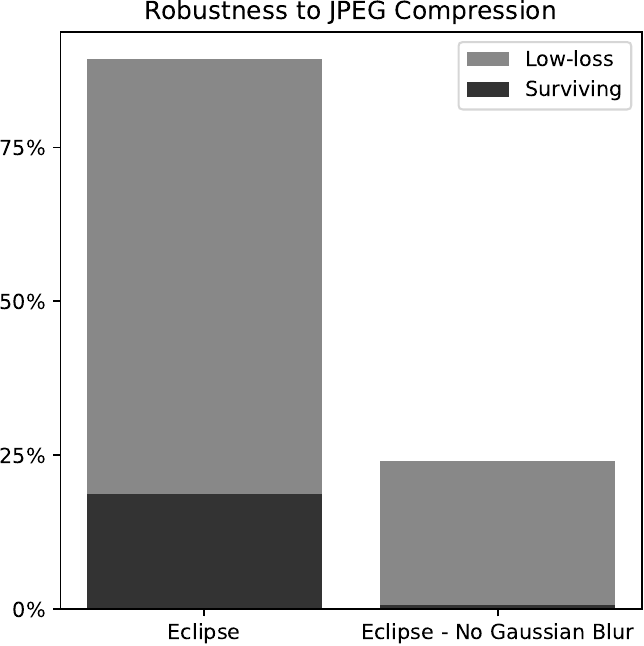}%
  \includegraphics[width=0.255\textwidth]{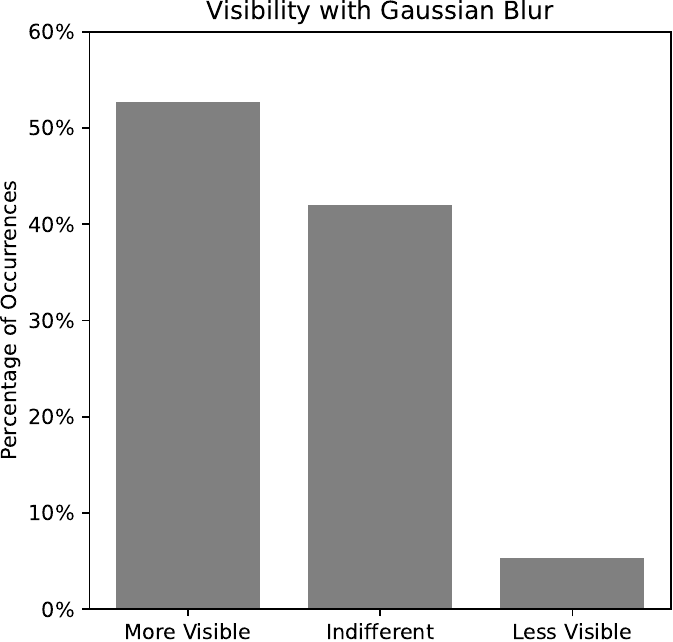}%
}\hfil
\subfloat[Local Surrogate.\label{fig:eclipse-ablation-local-surrogate}]{%
  \includegraphics[width=0.24\textwidth]{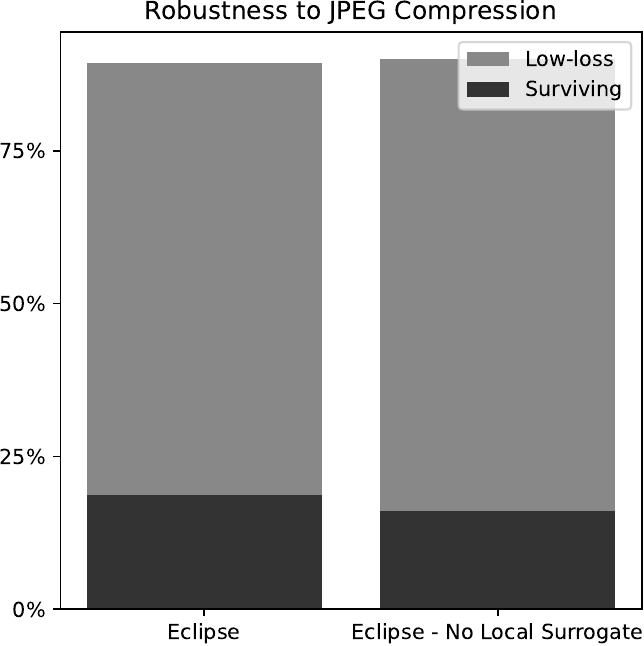}%
  \includegraphics[width=0.255\textwidth]{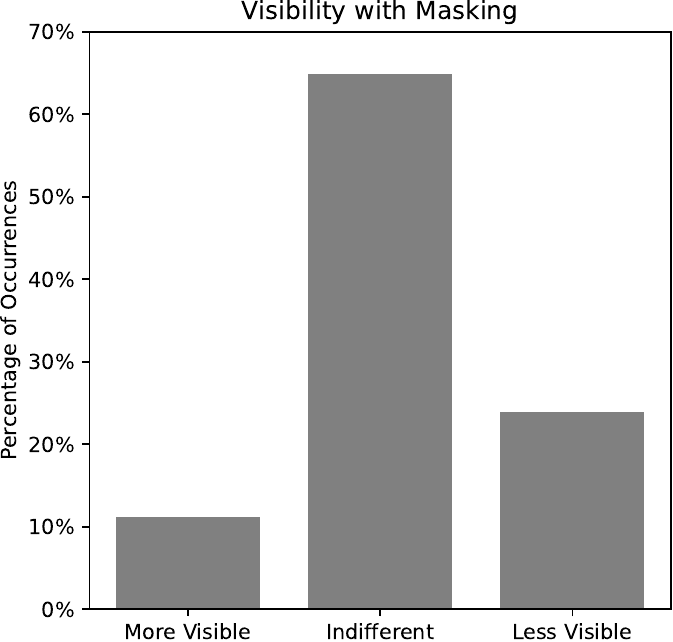}%
}

\caption{ECLIPSE ablation study results.}
\label{fig:eclipse-ablation}

\end{figure}

\mypar{ECLIPSE Ablation Study}
An ablation study was conducted to evaluate the impact of two novel components of the ECLIPSE attack: gradient Gaussian blurring and the sampling mask derived from the local surrogate. These components were assessed for their contribution to the three effectiveness properties. Robustness and detectability metrics mirror those used in attack comparisons, while visual stealthiness was evaluated via a survey. Participants were asked to compare ECLIPSE examples with ablated counterparts, choosing between three options: "The one on the left" (ECLIPSE), "The one on the right" (ablated version), or "I see no difference." Results illuminate the role of each component in the attack's overall effectiveness and stealth. Removing the Gaussian blur step significantly impacts all three effectiveness properties. Most notably, it improves robustness to processing, as evidenced by reduced loss and higher survival rates (Figure \ref{fig:eclipse-ablation-kernel}). Table \ref{tab:combined-eclipse-ablation} shows improved metrics, while stealthiness to automatic detection also benefits, with a $37\%$ reduction in the median number of queries and a $25\%$ decrease in the interquartile range (Table \ref{tab:combined-eclipse-ablation}). Despite a drop in the Area Under the Curve of the Receiver Operating Characteristic (ROC AUC) from 0.68 to 0.5, indicating minimal automatic detection performance without the blur, the inclusion of Gaussian blur does not improve visual stealthiness, with over $50\%$ of participants finding the adversarial examples more visible (Figure \ref{fig:eclipse-ablation-kernel}). On the other hand, removing the local surrogate primarily benefits stealthiness, with a drop in ROC AUC from 0.71 to 0.5 (Table \ref{tab:combined-eclipse-ablation}). It also reduces the number of queries by $18\%$, with a $10\%$ reduction in the interquartile range (Table \ref{tab:combined-eclipse-ablation}). However, it has minimal impact on robustness to processing or human visual stealth, with more than $60\%$ of participants perceiving no difference (Figure \ref{fig:eclipse-ablation-local-surrogate}) and negligible effects on JPEG compression metrics, indicating that Gaussian blur is the key component affecting these properties.

\begin{table}[bt]
    \centering
    \scriptsize
    \caption{Ablation study results of automatic detection of benign images and adversarial examples using spectral features. Each metric is a cross-validation score for Support Vector Machine classifiers. The are under the ROC curve is highlighted in bold.}
    \begin{tabular}{lrrrrr}
        \toprule
        Comparison & Accuracy & Precision & Recall & F1-score & \textbf{ROC AUC} \\
        \midrule
        ECLIPSE & 0.87 ($\pm$ 0.01) & 0.03 ($\pm$ 0.20) & 0.01 ($\pm$ 0.08) & 0.02 ($\pm$ 0.11) & \textbf{0.50 ($\pm$ 0.03)} \\
        No Gaussian blur & 0.90 ($\pm$ 0.03) & 0.69 ($\pm$ 0.23) & 0.39 ($\pm$ 0.18) & 0.50 ($\pm$ 0.18) & \textbf{0.68 ($\pm$ 0.09)} \\
        No Local Surrogate & 0.90 ($\pm$ 0.04) & 0.71 ($\pm$ 0.21) & 0.46 ($\pm$ 0.28) & 0.54 ($\pm$ 0.22) & \textbf{0.71 ($\pm$ 0.14)} \\
        \bottomrule
    \end{tabular}
    \label{tab:eclipse-ablation-spectral-classification}
\end{table}

\begin{table}[bt]
    \centering
    \scriptsize
    \setlength{\tabcolsep}{8pt} 
    \caption{Ablation study results. The table combines Robustness to Compression metrics and remote model query count.}
    \begin{tabular}{lccc|ccc}
        \mbox{} & \multicolumn{3}{c}{Compression Robustness} & \multicolumn{2}{c}{Remote Queries} \\
        \cmidrule(lr){2-4} \cmidrule(lr){5-6}
        & Median Loss & Low-loss (\%) & Surviving(\%) & Median & IQR \\
        \midrule
        ECLIPSE & 0.15 & 89.33 & 18.67 & 8125 & 11212.50 \\
        No Gaussian blur & 0.36 & 24.00 & 0.67 & 12870 & 14917.50 \\
        No Local Surrogate & 0.17 & 90.00 & 16.00 & 9880 & 12382.50 \\
        \bottomrule
    \end{tabular}
    \label{tab:combined-eclipse-ablation}
\end{table}

\begin{figure}[t]
    \centering
    \includegraphics[width=0.5\linewidth]{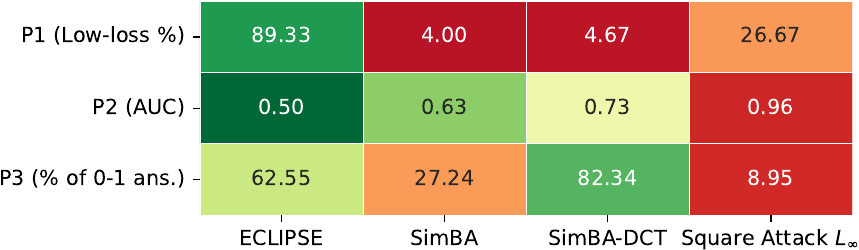}
    \caption{Summary of evaluation of the main metrics of each effectiveness property.}
    \label{fig:summary-heatmap}
\end{figure}

\mypar{Discussion} The results demonstrate that ECLIPSE achieves a superior balance across the three effectiveness properties. First, it generates the largest number of adversarial examples resilient to JPEG compression, substantially outperforming the second-best approach, the Square Attack. Second, ECLIPSE effectively bypasses defenses based on spectral features, as adversarial examples produced by the method are indistinguishable from benign images. Notably, a classifier trained for this distinction performs no better than random chance. While ECLIPSE does not set new benchmarks for query efficiency, its performance remains comparable to SimBA, one of the baseline methods. Furthermore, in terms of visual stealthiness, ECLIPSE achieves minimal degradation in human recognition rates relative to SimBA-DCT, the least detectable attack among those evaluated. A summary table of the main metrics that characterize the effectiveness properties for each attack is reported in Figure~\ref{fig:summary-heatmap}. To conclude, an ablation study of ECLIPSE’s novel components confirms their critical contributions to its effectiveness across assessed properties.
We shortly discuss the higher query count of ECLIPSE with respect to baselines. This is primarily due to the computational cost of gradient estimation via Hill Climbing, which exceeds that of heuristic baselines. However, as demonstrated in this section, this trade-off enhances real-world deployability. The increased query requirement imposes greater effort and cost on an attacker, necessitating multiple accounts and a more complex request distribution strategy. However, it does not drastically impact the real-world feasibility of the attack.

\section{Conclusions}
\label{sec:conclusions}
Adversarial examples present significant risks to machine learning systems, even in black-box scenarios where attackers rely on query-only access. While evasion attacks in computer vision are theoretically feasible, their practical deployment is constrained. We have formalized three \textit{effectiveness properties} to measure the real-world feasibility of an adversarial attack: Robustness to Compression, Stealthiness to Automatic Detection, and Stealthiness to Human Inspection. We have further presented ECLIPSE, an attack that balances these properties through two novel components: Gaussian blurring of estimated gradients and gradient masking using heatmaps derived from surrogate models. ECLIPSE demonstrates superior robustness to JPEG compression, achieving adversarial success in $89\%$ of cases compared to $27\%$ for Square Attack \(L_\infty\). Against spectral-based detection, ECLIPSE achieves perfect stealthiness (AUC $0.5$), significantly outperforming the Square Attack \(L_\infty\) (AUC $0.96$). In terms of visibility, ECLIPSE ranks second, closely behind SimBA-DCT, with $63\%$ of survey participants rating it negligible or invisible. Our research demonstrates the feasibility of adversarial attacks in real-world scenarios and highlights the necessity of developing defenses against more sophisticated threats. Through this evaluation, we demonstrated that existing State-of-the-Art attacks exhibit limited adherence to effectiveness properties, whereas ECLIPSE achieves a well-balanced trade-off.

\mypar{Future Work and Limitations}  
While the work considers few baselines compared to the vast literature of adversarial examples, it considers the most meaningful State-of-the-Art attacks. Future work could extend robustness evaluations to include additional image transformations and defensive measures, such as physical deployment through printed adversarial examples~\cite{wei_black-box_2021}. Moreover, exploring perturbations beyond additive noise, such as changes to brightness, contrast, or color dynamics, could enhance stealthiness under diverse conditions. A broader study incorporating a wider range of attacks would further refine the analysis of effectiveness properties and their trade-offs.

\begin{credits}

\mypar{\ackname}
This work was partially supported by the Google.org Impact Challenge - Tech for Social Good Research Grant (Tides Foundation) and project SERICS (PE00000014) under the MUR National Recovery and Resilience Plan funded by the European Union - NextGenerationEU.

\mypar{\discintname}
The authors have no competing interests to declare that are
relevant to the content of this article.

%
%
%

\mypar{Ethical Statement}
The survey on visual stealthiness did not collect any personal or identifiable information, including age, ensuring complete anonymity. Participation was voluntary and no incentive was given to participate. Even though no age-specific data was collected, we infer the survey demographic to be mostly under 30, given the channels where the survey was spread (student mailing list and social media).
\end{credits}

%
%
%
\bibliographystyle{splncs04}
\bibliography{all_bib}




\end{document}